\documentclass[aps,twocolumn,epsf]{revtex4}
\usepackage{graphics}
\usepackage{graphicx}
\usepackage{epsfig}

\begin{document}
\input epsf.tex
\bibliographystyle{apsrev}

\title{Catching Super Massive Black Hole Binaries Without a Net}
\author{Neil J. Cornish and Edward K. Porter}
\affiliation{Department of Physics, Montana State University, Bozeman, MT 59717}

\begin{abstract}
The gravitational wave signals from coalescing Supermassive Black Hole Binaries
are prime targets for the Laser Interferometer Space Antenna (LISA). With
optimal data processing techniques, the LISA observatory should be able to detect
black hole mergers anywhere in the Universe. The challenge is to find ways to
dig the signals out of a combination of instrument noise and the large foreground
from stellar mass binaries in our own galaxy. The standard procedure of matched filtering
against a grid of templates can be computationally prohibitive, especially when the
black holes are spinning or the mass ratio is large. Here we develop an alternative
approach based on Metropolis-Hastings sampling and simulated annealing that is orders
of magnitude cheaper than a grid search. We demonstrate our approach on simulated LISA data
streams that contain the signals from binary systems of Schwarzschild Black Holes, embedded in
instrument noise and a foreground containing 26 million galactic binaries. The
search algorithm is able to accurately recover the 9 parameters that describe the black hole
binary without first having to remove any of the bright foreground sources, even when
the black hole system has low signal-to-noise.
\end{abstract}
\pacs{}

\maketitle

Supermassive Black Hole Binaries (SMBHBs) and Extreme Mass Ratio Inspirals (EMRIs)
of a compact object into a supermassive black hole are two of the most exciting targets
for the LISA observatory~\cite{LISA}. Studies of these objects will yield
insights into the role played by black holes in structure formation and galactic dynamics.
The signals will also encode information about strong field, dynamical gravity that can
be used to perform precision tests of general relativity~\cite{ryan,hughes,berti}.

The SMBHB and EMRI signals contain a wealth of information that is encoded in a highly modulated
time series composed of multiple harmonics of several distinct, evolving periods. The complexity
of the signal is good news in terms of the science yield, but it poses a significant challenge
to the data analyst. The signal from a binary system of structureless spinning objects,
as described by  general relativity and detected by LISA, is controlled by 17 parameters. In the
case of SMBHBs the systems are expected to have circularized before entering the LISA band, thereby
reducing the search to 15 parameters. In the case of EMRIs the orbits are expected to maintain
significant eccentricity in the LISA band, but the spin of the smaller body can be neglected, thereby
reducing the search to 14 parameters.

The large dimension of the search spaces and the high computational cost of generating the search
templates make SMBHBs and EMRIs challenging targets for data analysis~\cite{gair}. The problem only
gets worse when one considers that we need to extract these signals from a timeseries that also
contains the signals from millions of galactic binaries, and in the case of EMRIs, a possible
self-confusion from hundreds of other EMRI systems~\cite{barackcutler2}.

It has been estimated that it would take $10^{40}$ templates to perform an optimal grid search
for EMRI signals~\cite{gair}. The numbers are less for SMBHBs, but still out of reach computationally.
Several alternative approaches have been discussed, including non-template based strategies
that look for tracks in spectrograms~\cite{wen}, and hierarchical, semi-coherent grid based
searches~\cite{gair}. Here we consider an alternative approach that uses Metropolis-Hastings
sampling and simulated annealing to search through the space of templates. Our search method is
closely related to the Markov Chain Monte Carlo (MCMC)~\cite{gilks,gamer} method that is used to
explore the posterior distribution of the model parameters once the source has been located.
In previous work the MCMC approach was used to test the Fisher matrix predictions for SMBHB
parameter uncertainties by
starting the chains off very close to true source parameters~\cite{wick,corpor}. It was found
that even the more sophisticated adaptive Reverse Jump MCMC algorithm performed poorly
when searching large regions of parameter space. We have found the non-Markovian sampling employed
by our algorithm to be many orders of magnitude faster than the MCMC search algorithms that
have been investigated to-date. Advanced MCMC techniques that employ importance resampling
and well designed priors have been used to study 5-parameter binary inspiral signals in the
context of ground based gravitational wave detectors~\cite{nelson}. It would be very interesting
to see how this algorithm performs in the LISA context. We apply our search algorithm to
simulated LISA data streams that include the signals from a pair of non-rotating black holes and a
foreground produced by galactic white dwarf binaries. While the SMBHB system we consider is simpler
than the general case (the model is described by 9 parameters rather than 15), it serves to illustrate
the relative economy of the gridless approach.

The gravitational waveform for a supermassive black hole system consisting of two Schwarzschild black
holes is described by 9 parameters: the redshifted chirp mass, $M_{c}$; the redshifted
reduced-mass, $\mu$; the sky location, $(\theta,\phi)$;  the time-to-coalescence, $t_{c}$;
the inclination of the orbit of the binary, $\iota$; the phase of the wave at coalescence, $\varphi_{c}$;
the luminosity distance, $D_{L}$; and the polarization angle, $\psi$.  The parameters $M_{c},\mu$ are
intrinsic to the system, while $D_{L}, \iota, \varphi_{c}, \psi$ are extrinsic as they depend on the
perspective of the observer. The other three parameters, $(\theta,\phi)$ and $t_c$, would be extrinsic
if the LISA observatory were static, but the motion of the detector couples these parameters to
the intrinsic evolution. The extrinsic parameters $D_{L}, \iota, \varphi_{c}, \psi$ can be analytically
solved for using a generalized F-statistic~\cite{JKS}, leaving a 5 dimensional search space.

\begin{figure}[t]
\begin{center}
\epsfig{file=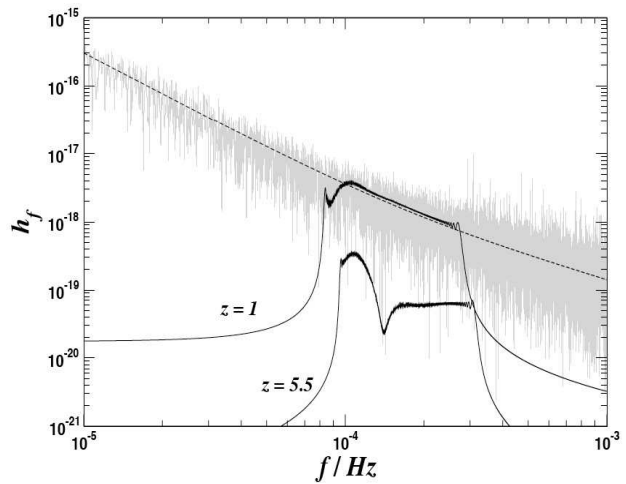, width=3in, height=2.3in}
\end{center}
\caption{The strain spectral density in a single LISA channel. The grey line is the LISA response to
a galactic background of ~26 million sources plus simulated instrumental noise. The solid black lines show the
LISA response to the SMBHB signals alone. The dashed black line indicates the RMS instrument plus galactic
confusion noise level.}
\label{fig:signal}
\end{figure}

We illustrate the performance of the search algorithm by considering two representative LISA sources -
a $10^{6}-10^{5}\,M_{\odot}$ binary system at $z=1$ and a $10^{5}- 5 \times 10^{4}\,M_{\odot}$ binary
system at $z=5.5$. In each case the time of observation is 6 months, and the observations end
$\sim 1$ week prior to merger. The early termination of the signal is designed to demonstrate
LISA's ability to give early warning to other telescope facilities. The $z=1$ example has parameters
$(M_{c}/M_{\odot},\mu/M_{\odot}, D_{L}/{\rm Gpc}, t_c/{\rm months}, \theta, \phi, \iota,
\varphi_{c}, \psi) = (4.93\times10^{5},
1.82\times10^{5}, 6.6, 6.23, 1.325, 2.04, 1.02, 0.95, 0.66)$
and the $z=5.5$ example has parameters 
$(M_{c}/M_{\odot},\mu/M_{\odot}, D_{L}/{\rm Gpc}, t_c/{\rm months}, \theta, \phi, \iota,
\varphi_{c}, \psi) = (3.95\times10^{5}, 2.17\times10^{5}, 53, 6.25, 1.927,
0.351, 1.318, 2.0, 0.23)$. To make the searches more realistic, we add in a galactic foreground
consisting of approximately 26 million galactic sources. The galactic binary foreground is generated using
a Nelemans, Yungelson and Zwart galaxy model~\cite{NYZ, TRC}. The signal-to-noise ratio (SNR) for the sources
is estimated using the combined instrument and galactic confusion noise. The $z=1$ example has
${\rm SNR} = 118.0$ and the $z=5.5$ example has ${\rm SNR} = 9.87$. These SNR ratios are on the low
side for typical LISA observations of SMBHBs as we terminate the observations a week before merger. The
full inspiral signals would give SNRs of $\sim 387$ and $\sim 182$ for the two cases, and the merger
and ringdown signals would further boost the SNRs by a factor of $\sim 2$ or more. In Fig~\ref{fig:signal} we plot
the detector response to the galactic foreground and instrument noise, along with the noise-free response
to the SMBHB signals.  We use restricted post-Newtonian waveforms with 2-PN evolution of the phase and
we employ the two independent interferometry channels that are available at low frequencies~\cite{cutler98}. 
As the equations that describe the phase evolution break down before we reach the last stable circular
orbit at $R=6M$, we terminate the search templates at a maximum value of $R=7M$. For the sources in question,
the observation period terminates a week from coalescence, so the maximum gravitational wave frequency
reached is $0.28$ mHz for the $z=1$ example and $0.32$ mHz for the $z=5.5$ example.  With this frequency
range, the SMBHBs overlap with over 22.5 million galactic binaries. To minimize the computational cost, the
search templates were generated at a sample cadence of 4.2 mHz.

Our search algorithm uses Metropolis-Hastings rejection sampling, simulated annealing and algebraic
extremization over extrinsic and quasi-extrinsic parameters. The sampling proceeds as follows: Choose a random
starting point $\vec{x}$ in parameter space.
Using a proposal distribution $q(\cdot \vert \vec{x})$, draw a new point $\vec{y}$. Evaluate the
Hastings ratio
\begin{equation}
H = \frac{\pi(\vec{y}) p(s \vert \vec{y}) q(\vec{x} \vert \vec{y})}
{\pi(\vec{x}) p(s \vert \vec{x}) q(\vec{y} \vert \vec{x})} \, .
\end{equation}
Accept the candidate point $\vec{y}$ with probability $\alpha = {\rm min}(1,H)$, otherwise remain
at the current state $\vec{x}$. Here $\pi(\vec{x})$ are the priors on the parameters,
\begin{equation}
p(s|\vec{x}) = {\rm const.} \; e^{-\langle s-h(\vec{x})|s-h(\vec{x})\rangle/2},  
\end{equation}
is the likelihood and $q(\vec{x}|\vec{y})$ is the proposal distribution.  The angular brackets
$\langle s-h(\vec{x})|s-h(\vec{x})\rangle$ denote the standard noise weighted inner product of the
signal $s$ minus the template $h(\vec{x})$. We employ three different proposal distributions that
are designed to give small, medium and large jumps. This mixture of jump sizes gives the search the
flexibility to fully explore the parameter space and the ability to quickly hone in on promising regions.
The small jumps are drawn from a
multi-variate Normal distribution, the medium sized jumps are given by a uniform draw of $\pm 10 \sigma$
in each parameter and the large jumps come from a full range, uniform draw on all the parameters.
We used a mixture of 20 small jump proposals for every medium or large jump proposal. 
Correlations between the parameters can seriously hurt the acceptance rate,
so we use a multi-variate Normal distribution that is the product of Normal distributions in each
eigendirection of the Fisher information matrix, $\Gamma_{ij}(\vec{x})$.  The standard deviation in
each eigendirection is set equal to $\sigma_{i} = 1/\sqrt{DE_{i}}$, where $D=9$ and $E_{i}$ is the
corresponding eigenvalue of the Fisher matrix~\cite{corcro}. The Fisher matrix is also used to
scale the medium size jumps.

The simulated annealing is done by multiplying the noise weighted inner product $\langle s \vert h\rangle$
by an inverse temperature $\beta$. We used a standard power-law cooling schedule:
\begin{equation}
\beta = \left\{ 
\begin{array}{ll} 10^{B(1-i/N_{c})} & 0\leq i\leq N_{c} \\ \\ 1
& i > N_{c}  
\end{array}\right.,
\end{equation}
where $i$ is the number of steps in the chain and $N_{c}$ is the number of steps the chain takes to
reach the normal temperature. We found that an initial heat factor of between 10 to 100 and a cooling schedule
that lasted for $\sim 10^{4}$ steps worked well, but the performance was not particularly sensitive to these
choices. For low SNR sources smaller initial heat factors and slightly longer cooling schedules yielded
better results.

The F-statistic is used to automatically extremize over the four parameters $(D_L, \iota,\psi,\varphi_c)$, but
the motion of the LISA detector sets a time reference, so the usual trick of using a fast Fourier transform
to extremize over the time to coalescence, $t_c$, is not strictly permitted. However, the waveforms are
much less sensitive to the sky location than they are to $t_c$, so we employed $t_c$ maximization during
the annealing phase for the large and medium jump proposals. This procedure biases the solution, but the
bias is erased by subsequent jumps.

In dozens of tests applied to many different examples, our search
algorithm never failed to detect the SMBHB signals. On occasions the chain would lock onto
a secondary maxima of the likelihood function, but this behaviour can be heavily suppressed by
using longer cooling schedules. Once the annealing phase is complete the $t_c$ maximization
is turned off and our search algorithm becomes a standard Markov Chain Monte Carlo
(MCMC) algorithm for exploring the posterior distribution function.
The MCMC method is a multi-purpose approach that can be used to perform model comparisons,
estimate instrument noise, and provide error estimates for the recovered parameters~\cite{gilks,gamer}.
The method is now in widespread use in many fields, and
is starting to be used by astronomers and cosmologists. MCMC techniques have been applied
to ground based gravitational wave data analysis~\cite{christ}; a toy LISA problem~\cite{umstat};
and the extraction of multiple overlapping galactic binaries from simulated LISA data~\cite{corcro}.

For the example at $z=1$ we use the following uniform priors in our search: we
choose the mass ratio to lie between 5 and 15, the redshifted total mass between $5\times10^{5}$ and
$5\times10^{6}$ solar masses, $t_{c}$ is chosen to lie within 3 and 9 months, and $\theta$ and $\phi$
are drawn from a uniform sky distribution.
The initial heat was set at 100 ($B=-2$) and the annealing lasted for $N_c=10,000$ steps. The search took three
hours to run on a single 2 GHz processor.
\begin{figure}[t]
\begin{center}
\epsfig{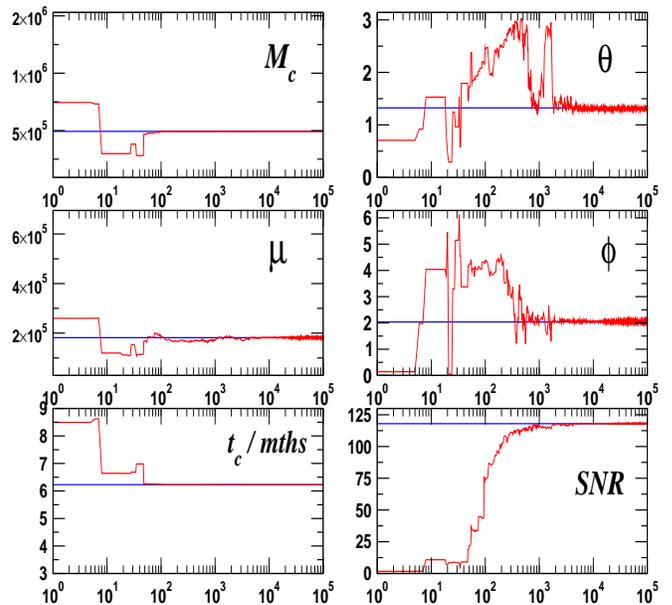}
\end{center}
\caption{A plot of the search chains for the five intrinsic parameters and the SNR. 
In all cases, the straight solid line represents the true values of the SMBHB parameters.
After $N=N_c=10000$ steps the search becomes a standard MCMC exploration of the posterior
distribution function.}
\label{fig:chns}
\end{figure}
In Fig.~\ref{fig:chns} we plot a representative search chain. Because the search algorithm
locks onto the source in $N \sim 1000$ steps, we use a logarithmic scale for the number of
iterations, $N$. The $t_c$ maximization allows the search to hone in on $M_{c}$ and $t_{c}$ very quickly.
The reduced mass $\mu$ is less well constrained and takes a little longer to lock in, and the
sky location gets fixed last of all. The extrinsic parameters $D_L,\iota$ are
recovered once the sky location is determined, while $\psi$ and $\varphi_c$ continue to explore
their full range throughout the evolution. The failure to fix $\psi$ and $\varphi_c$ is consistent
with the Fisher matrix predictions for the uncertainties in these parameters. The errors
in the recovered search parameters were: $\Delta M_c = -42 M_\odot =-0.154 \sigma$; $\Delta \mu = 729 M_\odot =
0.147 \sigma$; $\Delta t_c = 76 \, {\rm s} = -0.171 \sigma$;
$\Delta \theta = 0.82^\circ = 1.06 \sigma$; and $\Delta \phi = 0.65^\circ = -1.28 \sigma$;
where the standard deviations were determined from the MCMC portion of the chains.

We found that our gridless search algorithm is able to reliably identify the SMBHB signal within
$\sim 1,000$ steps. We have calculated that it would take $9.3\times 10^{12}$ templates to cover the same
search range with an F-statistic based grid search at a minimal match level of 0.9~\cite{owen}.
The comparison is not entirely fair since we also used an illegal maximization over $t_c$ during
the annealing phase, but we have verified that the annealed chains are able to find the SMBHB
signal without this trick, it just takes 10 to 100 times longer. Either way, our search algorithm
is significantly more economical than a naive grid based search.

\begin{figure}[t]
\begin{center}
\epsfig{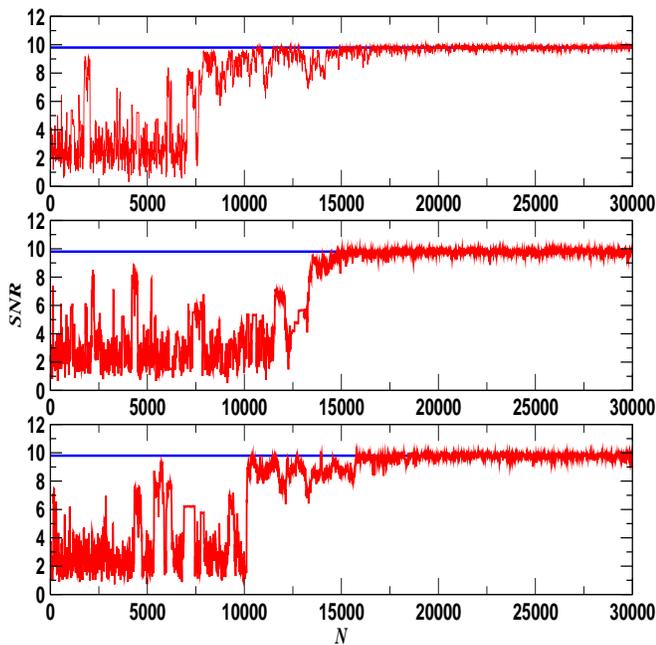}
\end{center}
\caption{Examples of the SNR evolution for three runs searching for the low SNR signal at
redshift $z=5.5$. The chains typically found, then lost the signal on several occasions early in
the runs due to the high temperature and low SNR. The chains usually locked on for good at
around iteration $N\sim 12000$.}
\label{fig:low}
\end{figure}

For the example at $z=5.5$ we use the following uniform priors in our search: we
choose the mass ratio to lie between 1 and 5, the redshifted total mass between $2\times10^{5}$ and
$2\times10^{6}$ solar masses, $t_{c}$ is chosen to lie within 5 and 7 months and $\theta$ and $\phi$
are drawn from a uniform sky distribution. The initial heat was set at 10 $(B=-1)$
and the annealing phase lasted for $N_c=20,000$ steps. In Fig.~\ref{fig:low} we plot the SNR evolution
for three runs. Two of these runs happened to lock
onto an alternative solution for the sky location that exist because of the approximate
symmetry $\phi \rightarrow \phi + \pi$ and $\theta \rightarrow \pi - \theta$ that holds for
the low frequency LISA response function. Since the two solutions for the sky position have
almost equal likelihood, the bimodality of the solution is a feature, rather than a flaw, of the
search algorithm. As might be expected, the search algorithm takes longer to lock onto weak sources
than strong sources, but the run times are still measured in hours, not days. 

Here we have shown that it is possible to dig a SMBHB signal out from under instrument noise and the signals from
foreground sources. The errors in the recovered parameters are consistent with a Fisher matrix prediction
that treats the galactic foreground as an addition source of Gaussian noise. We will present a detailed
study of detection threshold and the posterior distributions in the presence of galactic foregrounds
in a future publication.
The next step is to apply the same techniques to the more complicated signals from spinning SMBHB's
and EMRIs. The larger parameter spaces are not expected to pose a problem as the search cost
is expected to scale linearly with the search dimension. Indeed, it should be possible to
simultaneously search for multiple, overlapping EMRI signals. We consider our current work as a
proof-of-principle that the LISA data analysis challenge can be addressed with modest computational
resources.

\end{document}